\documentclass{elsart}

\usepackage{graphicx}
\usepackage{epsfig,array}
\usepackage{booktabs}

\newcommand{\ket}[1]{\left| {#1} \right\rangle}
\newcommand{\bra}[1]{\left\langle {#1}\right|}

\newcommand{\beq}{\begin{equation}}
\newcommand{\eeq}{\end{equation}}
\newcommand{\beqa}{\begin{eqnarray}}
\newcommand{\eeqa}{\end{eqnarray}}

\newcommand{\sla}[1]%
        {\kern .25em\raise.18ex\hbox{$/$}\kern-.75em #1}
\newcommand{\mybar}[1]%
        {\kern 0.8pt\overline{\kern -0.8pt#1\kern -0.8pt}\kern 0.8pt}

\newcommand{\veb}[1]{{\vec{#1}}}
\newcommand{\mo}[1]{\left\| {#1} \right\|}

\begin{document} 

\begin{frontmatter}

\title{Non leptonic two--body decay amplitudes from finite volume calculations}

\author[romeII]{Giulia~M.~de~Divitiis}
\author[romeII]{Nazario~Tantalo}

\address[romeII]{University of Rome ``Tor Vergata'' and INFN sez. RomaII, 
	Via della Ricerca Scientifica 1, 
	I-00133 Rome}

\begin{abstract}
We discuss the quantization of the energy levels of two--particle
scattering states in a finite volume in the case of Bloch--type boundary conditions.
A generalization of the L\"uscher quantization condition is obtained 
that can be used in order to calculate
the scattering phases,
resulting for example by the strong elastic interaction of two pions, 
at fixed physical momentum transfers on a sequence of volumes of growing sizes.
We also give a generalization of the Lellouch--L\"uscher formula to be used
to extract the physical $K\mapsto \pi\pi$ decay rate below the inelastic threshold
from finite volume calculations. The formula is valid up to corrections
exponentially vanishing in the volume. 
By using this formula the calculation can in principle be performed on different
finite volumes of growing sizes in order to keep under control the corrections. 
\end{abstract}

\end{frontmatter}

\section{Introduction}
\label{sec:introduction} 

The calculation of the weak matrix elements associated to the non--leptonic
decay $K\mapsto \pi\pi$ requires a non--perturbative level of accuracy, due to the strong
nature of the final state re--interactions.
Lattice gauge theory is the best candidate among the other methods
to achieve a phenomenologically relevant result for this process but there are theoretical problems
that have to be addressed before such kind of calculations can be actually
performed.
The obstacles that have to be removed are mainly due to two different reasons.

The so called ``ultraviolet problem'', concerning the construction of finite matrix
elements of properly renormalized lattice operators, has been addressed
by many authors and we will not discuss it further in the rest of this paper
(see~\cite{Becirevic:2004fw} for a recent review of this subject). Here we will focus 
our attention on the so called ``infrared problem''.

The infrared problem, formalized in the well known Maiani--Testa ``no--go'' 
theorem~\cite{Maiani:1990ca,Ciuchini:1996mq}, concerns the impossibility
of connecting the large time correlation functions calculated on a large euclidean volume
with the physical decay amplitudes.
A solution to the infrared problem has been found in ref.~\cite{Lellouch:2000pv} 
where Lellouch and L\"uscher (in the following referred as LL) derived a relation
connecting the finite volume euclidean $K\mapsto \pi\pi$ matrix elements with the
physical ones, up to exponentially small finite volume corrections.

The key point in the derivation of the LL formula is that the Maiani--Testa
theorem applies in the case of very large volumes, where the spectrum of the
two--particle final state is continuous, but ceases to be valid in the case 
of finite volumes where the two--particle energy is quantized. In the last case
a kaon at rest cannot decay into two pions unless the physical extension
of the finite volume is such that one of the two--particle
energy levels equals the mass of the decaying particle.
A simple phenomenological analysis based on the results of
refs.~\cite{Luscher:1985dn,Luscher:1986pf,Luscher:1990ux,Luscher:1991cf} 
shows that, when the  two pions satisfy periodic
boundary conditions,
the volume on which the energy of the first excited state of definite isospin
coincides with the kaon mass is of the order of $5.5$~fm.
Such a large value of the volume makes the calculation unfeasible
from the practical point of view due to the limiting present computer power.
Furthermore the calculation is performed on a single finite volume and
the corrections, although exponentially vanishing, cannot be quantified.

The authors of ref.~\cite{Lin:2001ek} have shown that a set of LL formulas can be derived
also when the energy of the outgoing two--pion state does not coincide with the
kaon mass. In this case the resulting infinite volume decay rates are not calculated
at the physical point, i.e. at the values of the two--pion energies fixed
by relativistic kinematics. Nevertheless one can repeat the calculation for
different finite volumes and then extrapolate the results to the physical
point. In ref.~\cite{Lin:2001ek} it has been also pointed out that
the finite volume corrections to the LL formulas could result to be
sizable if only few two--particle states have energies below the
inelastic threshold.

In this paper we study the quantization condition of the energy of a two--particle state
on a finite volume by using a generalized set of boundary conditions
($\theta$--BC) for the two particles.
In ref.~\cite{deDivitiis:2004kq} we have shown  that the use of $\theta$--BC in
framework of lattice calculations makes possible a continuous
momentum transfer between one--particle states.
In the following we obtain a quantization condition for the energy of a two--particle state
on a finite volume that generalize a result previously
obtained by L\"uscher~\cite{Luscher:1990ux} in the case of periodic boundary conditions.
We also give a generalization of the LL formula and
show that by using the $\theta$--dependence of this formula and of our
quantization condition it is possible to find a sequence of
finite volumes of growing sizes on which
the calculation of the $K\mapsto \pi\pi$ decay rates can be performed at the physical
point. Using the values of the scattering phases predicted
by chiral perturbation theory we show that the physical amplitudes can be calculated on volumes
of the order of about $3$~fm and, at the same time, the residual functional dependence 
of the results upon the volume can be quantified.

The plan of the paper is as follows. In section~\ref{sec:2pinafinitevolume}
we set up the notations. In section~\ref{sec:lessons} we derive
the quantization condition while the generalized LL formula is given
in section~\ref{sec:genLLformula}. In section~\ref{sec:voldep}
we study the volume dependence of the two--pion spectrum
for some particular choices of the $\theta$--angle and
in section~\ref{sec:conc} we draw our conclusions.
Some technical details needed for the derivation of the quantization condition
are given in the appendices.

\section{Two--particle states in a finite volume}
\label{sec:2pinafinitevolume} 

The spectrum of a two--particle state on a finite volume in quantum field theory 
has been already studied in great detail
in refs.~\cite{Luscher:1985dn,Luscher:1986pf,Luscher:1990ux,Luscher:1991cf}
in the case in which the two particles satisfy periodic boundary conditions.
An energy quantization condition has been found by establishing in ref.~\cite{Luscher:1985dn,Luscher:1990ck}
a connection between quantum field theory and non--relativistic quantum mechanics.
Indeed, by assuming that
the two particles are spinless bosons of equal mass $m$ whose
dynamics can be described by a scalar field theory of the $\phi^4$--type,
that the reflection symmetry $\phi\mapsto -\phi$ is unbroken
and that the one--particle states are odd under this symmetry,
an effective Schr\"odinger equation can be written for the two--particle
state. In the center-of-mass reference frame this equation reads
\begin{equation}
-\frac{1}{2\mu}\triangle \psi(\veb{x}) +\frac{1}{2}\int{d\veb{x}^\prime\ U_E(\veb{x},\veb{x}^\prime)\
\psi(\veb{x}^\prime)} = E \psi(\veb{x})
\label{eq:effscreq}
\end{equation}
where the parameter $E$ does not represent the true energy
of the system, that we call $\mathcal{E}$, but it is connected to the last
through
\begin{equation}
\mathcal{E} = 2\sqrt{m^2+mE}
\end{equation}
In eq.~(\ref{eq:effscreq}) the parameter $\mu$ represent the reduced mass of the two--particle system while
$U_E(\veb{x},\veb{x}^\prime)$ is the Fourier transform
of the modified Bethe--Salpeter kernel~$\hat{U}_E(\veb{k},\veb{k}^\prime)$ 
introduced in~\cite{Luscher:1985dn}.
The ``pseudo--potential'' $U_E(\veb{x},\veb{x}^\prime)$ 
depends analytically on $E$ in the range $-m<E<3m$,
is a smooth function of the
coordinates $\veb{x}$ and $\veb{x}^\prime$ decaying exponentially in each direction
and is rotationally invariant so that one can pass to the radial effective Schr\"odinger
equation.

Thanks to these observations, in the following we will carry on the calculation of the spectrum of
a two--particle state in a finite volume by using a purely non--relativistic Hamiltonian that,
separating the center of mass motion from the internal motion,
comes out to depend upon the relative coordinate $\veb{r} = \veb{x}-\veb{y}$ only
\begin{equation}
\hat{H} = -\frac{1}{2\mu}\triangle + V(r), \qquad r = \left\|\veb{r}\right\|
\label{eq:crystham}
\end{equation}
where $\triangle$ is the Laplacian operator with respect to $\veb{r}$.
The potential is assumed to be spherically symmetric, a smooth function
of its argument and of finite range, i.e.
\begin{equation}
V(r) = 0 \qquad \mbox{for} \qquad r > R
\label{eq:finiterange}
\end{equation}
We will solve the problem on a finite cubical box of linear extension $L$
in each direction greater than the potential radius ($L>R$) 
and we will assume that the potential is periodic of period $L$
\begin{equation}
V(\mo{\veb{r}+\veb{n}L}) =  V(r)\qquad \mbox{for} \qquad \veb{n} \in {\bf Z}^3 
\label{eq:potperiodic}
\end{equation}
We can imagine to start from a given finite--range potential $\mathcal{V}(r)$ that
describe the interactions of the two particles and to build a periodic potential
as follows
\begin{equation}
V(\veb{r}) = \sum_{\veb{n} \in {\bf Z}^3}{
\mathcal{V}(\mo{\veb{r}+\veb{n}L})}
\label{eq:ourpotential}
\end{equation}
By construction $V(\veb{r})$ satisfies the periodicity condition stated in eq.~(\ref{eq:potperiodic}).
 
There are two differences between the potential $V(\veb{r})$ that we are going
to study and the pseudo--potential $U_E(\veb{x},\veb{x}^\prime)$.
The first one concerns the energy dependence of $U_E$ but this does not represent
a problem because all the results we are going to derive will be obtained at
fixed $E$.
The second difference between the quantum field system and the non--relativistic
one is that $U_E(\veb{x},\veb{x}^\prime)$ does not vanish if one between $\veb{x}$
and $\veb{x}^\prime$ is greater than $R$ but has exponentially small corrections.
Furthermore in the quantum field system there are additional 
exponentially small finite volume corrections that
arise from polarization effects.
For these reasons the results we are going to derive in the non--relativistic theory
will be valid also in the relativistic theory
up to exponentially small corrections.

The matching with a quantum mechanical system could be avoided at all following
an approach similar to that developed in ref.~\cite{Lin:2001ek}. In the following
we will not follow this strategy because, as will be clear later on in the
derivation, the quantum mechanical analogy allows us to benefit without additional effort
of a series of theoretical results obtained in the framework of solid
state physics and useful also in our case.

\section{Quantization condition: lessons from solid state physics}
\label{sec:lessons} 

In this section we derive a powerful relation connecting
the energy eigenvalues of a two--particle state on a finite volume
with the infinite volume scattering phases of the two particles.
By using the result stated in the previous section we will perform
the calculation in NRQM being the results valid also in QFT
up to exponentially vanishing finite volume corrections.

In order to obtain the energy quantization condition we 
have to deal with the Schr\"odinger equation for a particle in a periodic
potential, i.e. the same equation satisfied by electrons, holes
and excitons in a periodic crystal.  
We can thus export useful results well known to the solid state
physics community since a long time, provided that we interpret the
``cell size'' of the crystal as the physical extension of
the finite volume, i.e. $L$.

\subsection{Bloch's theorem}

In order to better understand the role of our particular
choice of the boundary conditions we want to recall the 
well known Bloch's theorem
\vskip 15pt
\newtheorem{theorem}{theorem}
\begin{theorem}
The wavefunctions of the ``crystal Hamiltonian'' can be written as the
product of a plane wave of wavevector $\veb{k}$ within the first Brillouin
zone, times an appropriate periodic function:
\begin{equation}
\psi_{\veb{k}}(\veb{r}) = e^{i\veb{k}\cdot\veb{r}}\ u_{\veb{k}}(\veb{r})
\label{eq:blochtheorem}
\end{equation}
where
\begin{equation}
u_{\veb{k}}(\veb{r}+\veb{n}L) = u_{\veb{k}}(\veb{r}) \qquad \mbox{and} \qquad
0 \le k_i < \frac{2\pi}{L} 
\label{eq:blochtheorem2}
\end{equation}
\label{teor:bloch}
\end{theorem}
Let us observe that we are dealing exactly with a sort of crystal Hamiltonian
(see eqs.~(\ref{eq:crystham}),~(\ref{eq:finiterange}) and~(\ref{eq:potperiodic}))
and so that the result stated from the Bloch's theorem apply straightforwardly
also in our case.
Furthermore, we observe that the wavefunctions $\psi_{\veb{k}}(\veb{r})$ do not satisfy
periodic boundary conditions but the more general set of b.c.
\begin{equation}
\psi_{\veb{k}}(\veb{r}+\veb{n}L) = e^{i\veb{k}\cdot\veb{n}L}\ \psi_{\veb{k}}(\veb{r})
\label{eq:blochcond}
\end{equation}
These boundary conditions have been recently considered in the framework
of lattice QCD in ref.~\cite{deDivitiis:2004kq} under the name
of $\theta$--boundary conditions, precisely in the form
\begin{equation}
\psi_{\veb{\theta}}(\veb{r}+\veb{n}L) = e^{i\veb{\theta}\cdot\veb{n}}\ \psi_{\veb{\theta}}(\veb{r})
\end{equation}
in order to handle in lattice simulations physical momenta smaller
than the allowed
\begin{equation}
\veb{p} = \frac{2\pi}{L}\veb{n}\;, \qquad \veb{n} \in {\bf Z}^3
\end{equation}
in the case of standard periodic boundary conditions. The matching between the
formalism of ref.~\cite{deDivitiis:2004kq} and the condition stating the fundamental
result of the Bloch's theorem is obtained by identifying
\begin{equation}
\veb{k} = \frac{\veb{\theta}}{L}
\end{equation}
%

\subsection{Korringa--Kohn--Rostoker theory}

Another fundamental result that we can gain from
solid state physics is the computational framework developed
independently by Korringa~\cite{Korringa:1947}, Kohn and Rostoker~\cite{Kohn:1954}
and known as the KKR method or the Green's function
method.
A straightforward application of this method will allow us to
derive the two--particle state energy quantization condition
in a simple and (in our opinion) clear way.

The KKR method can be applied under the hypotheses of a so called 
``muffin thin potential'', i.e.
a periodic, spherical symmetric potential that vanishes after a given
distance $R$ for each cell of the crystal.
After having realized that all these hypotheses are satisfied by our
potential defined in eq.~(\ref{eq:ourpotential}),
we can start in reviewing the KKR procedure
by considering the time--independent form of the Schr\"odinger equation
\begin{equation}
\left( \triangle + q^2 \right) \psi_{\veb{k}}(\veb{r}) = V^\prime(\veb{r})  \psi_{\veb{k}}(\veb{r})
\label{eq:nonhom}
\end{equation}
where we have defined
\begin{equation}
q^2 = 2\mu E
\end{equation}
and we have substituted $2\mu V(\veb{r})$ with $V^\prime(\veb{r})$.
In order to have a formal solution of this non--homogeneous partial
differential equation we introduce the free--particle Green's function
as the solution of
\begin{equation}
\left( \triangle_\veb{r} + q^2 \right) g(\veb{r}-\veb{r}_0;q) = \delta(\veb{r}-\veb{r}_0)
\end{equation}
Using the Green's function, the solution of eq.~(\ref{eq:nonhom}) can be written formally
as
\begin{equation}
\psi_{\veb{k}}(\veb{r}) = \phi_\veb{k}(\veb{r}) + 
\int_{-\infty}^{+\infty}{ d\veb{r}_0\ g(\veb{r}-\veb{r}_0;q)
V^\prime(\veb{r}_0)  \psi_{\veb{k}}(\veb{r}_0)
}
\label{eq:nonhom_sol}
\end{equation}
where $\phi_\veb{k}(\veb{r})$ is a solution of the homogeneous equation associated with 
eq.~(\ref{eq:nonhom}) with the additional requirement to satisfy the Bloch's condition
of eq.~(\ref{eq:blochcond}). 
In the previous equation the integration variable $\veb{r}_0$ spans the whole three dimensional
space and not only a period. 
We will set for the moment $\phi_\veb{k}(\veb{r}) = 0$
and we will come back to the case in which the homogeneous solution is present
later on.
We rewrite eq.~(\ref{eq:nonhom_sol}) as
\begin{equation}
\psi_{\veb{k}}(\veb{r}) = 
\int_{-\infty}^{+\infty}{ d\veb{r}_0\ g(\veb{r}-\veb{r}_0;q)
V^\prime(\veb{r}_0)  \psi_{\veb{k}}(\veb{r}_0)
}
\label{eq:nonhomsolnophi}
\end{equation}
We want to observe that no particular conditions are requested to 
the Green's function in order $\psi_{\veb{k}}(\veb{r})$ to satisfy
the Bloch's condition; indeed being $\psi_{\veb{k}}(\veb{r})$ present
in both the members of the previous equation and being the potential
periodic, the condition of eq.~(\ref{eq:blochcond}) is self--consistently
satisfied. For this reason we will not require $g(\veb{r}-\veb{r}_0;q)$
to satisfy any particular periodicity condition. As a consequence
we can use for this function the well known result
\begin{equation}
g(\veb{r}-\veb{r}_0;q) = -\frac{1}{4\pi} \frac{e^{iq\mo{\veb{r}-\veb{r}_0}}}{\mo{\veb{r}-\veb{r}_0}}
\end{equation}

The domain of integration in eq.~(\ref{eq:nonhomsolnophi})
can be reduced from the entire world to a single periodicity cell by introducing the ``greenian''
of the equation defined as
\begin{equation}
g_\veb{k}(\veb{r}-\veb{r}_0;q) = \sum_{\veb{n} \in {\bf Z}^3}{
e^{i\veb{k}\cdot\veb{n}L} g(\veb{r}-\veb{r}_0 - \veb{n}L;q)
}
\label{eq:greenian}
\end{equation}
Using the greenian definition together with the Bloch's condition and the periodicity
of our potential, the formal solution of the Schr\"odinger equation
can be rewritten as an integral spanning only a period
\begin{equation}
\psi_{\veb{k}}(\veb{r}) = 
\int_{\mbox{period}}{ d\veb{r}_0\ g_\veb{k}(\veb{r}-\veb{r}_0;q)
V^\prime(\veb{r}_0)  \psi_{\veb{k}}(\veb{r}_0)
}
\label{eq:nonhomsolnophi_cell}
\end{equation}
but, being the potential identically zero for distances greater than $R$, the
integration domain can be further reduced to a sphere of radius $R$
\begin{equation}
\psi_{\veb{k}}(\veb{r}) = 
\int_{S_R}{ d\veb{r}_0\ g_\veb{k}(\veb{r}-\veb{r}_0;q)
V^\prime(\veb{r}_0)  \psi_{\veb{k}}(\veb{r}_0)
}
\label{eq:nonhomsolnophi_R}
\end{equation}
Now we use the fact that, thank to the Schr\"odinger equation~(\ref{eq:nonhom}),
the previous relation can be rewritten in a form suitable for the application of
the Green's theorem
\begin{equation}
\psi_{\veb{k}}(\veb{r}) = 
\int_{S_R}{ d\veb{r}_0\ g_\veb{k}(\veb{r}-\veb{r}_0;q)
(\triangle_{\veb{r}_0} + q^2)  \psi_{\veb{k}}(\veb{r}_0)
}
\label{eq:nonhomsolnophiR_preG}
\end{equation}
Using the identity
\begin{equation}
g \triangle \psi = \psi \triangle g + \veb{\nabla} \cdot ( g \veb{\nabla} \psi -  \psi \veb{\nabla} g)
\end{equation}
and the Green's theorem we end up with a vanishing surface integral
\begin{equation} 
\int_{\partial S_R}{ dS_0\ 
\left[
g_\veb{k}(\veb{r}-\veb{r}_0;q) \frac{\partial \psi_{\veb{k}}(\veb{r}_0)}{\partial r_0}
-
\psi_{\veb{k}}(\veb{r}_0) \frac{\partial g_\veb{k}(\veb{r}-\veb{r}_0;q)}{\partial r_0}
\right]_{r_0=R}
} = 0
\label{eq:nonhomsolnophiR_G}
\end{equation}
The previous equation is the quantization condition we where
searching for. In the following, by using the expansion in spherical harmonics
of both the greenian and of the wavefunction,
we will recast this condition
in a system of equations expressing the 
two--particle scattering phases as functions of the energy eigenvalues and
vice versa.

\subsection{Partial wave expansion of the wavefunction}

Let us  consider the first periodicity cell. 
For distances greater than the potential radius,
the wavefunctions $\psi_{\veb{k}}(\veb{r})$ satisfy the free
particle equation and can thus be written as
\begin{equation}
\psi_{\veb{k}}(\veb{r}) = \sum_{lm}{
c_{lm}(\veb{k}) R_l(r;q) Y_{lm}(\hat{r}_0)
}
\label{eq:psiexp}
\end{equation}
where $c_{lm}(\veb{k})$ are coefficients to be determined by using
eq.~(\ref{eq:nonhomsolnophiR_G}) and $Y_{lm}(\theta,\phi)$ are the
spherical harmonics.
The radial part $R_l(r,q)$ can be expressed as
\begin{equation}
R_l(r,q) = \cos{ \delta_l(q)}\ j_l(qr) -
\sin{ \delta_l(q)}\ n_l(qr)
\qquad r \ge R 
\end{equation}
where $\delta_l(q)$ are the two--particle infinite volume scattering phases,
$j_l(qr)$ are the spherical Bessel functions and $n_l(qr)$ are
the spherical Neumann functions.
For later use we report the Wronskian relations satisfied by the
Bessel and Neumann functions
\begin{equation}
\left[j_l,R_l\right] = - \frac{\sin{\delta_l(q)}}{qr^2} \qquad
\left[n_l,R_l\right] = - \frac{\cos{\delta_l(q)}}{qr^2}
\label{eq:wronskian}
\end{equation}
where, as usual, the Wronskian of two functions is defined as
\begin{equation}
\left[f(x),g(x)\right] = f(x) g^\prime(x) - f^\prime(x)g(x)  
\end{equation}
%

\subsection{Partial wave expansion of the greenian}

The derivation of the partial wave expansion of the greenian
is a rather involved mathematical exercise. In order to
make our derivation of the quantization
condition of the two--particle energy as clear as possible 
we give all the technical details
in the appendix~\ref{sec:app1} and report here below only the
resulting expression
\begin{eqnarray}
g_\veb{k}(\veb{r}-\veb{r}_0;q) &=&
q\sum_{lm}{
j_l(qr)\ Y_{lm}(\hat{r})
n_l(qr_0)\ Y_{lm}^*(\hat{r}_0)
}
\nonumber \\ \nonumber \\
&+&
\sum_{lm l^\prime m^\prime}{
j_l(qr)\ Y_{lm}(\hat{r})\
\Gamma_{lm,l^\prime m^\prime}(\veb{k},q)\
j_{l^\prime}(qr_0)\ Y_{l^\prime m^\prime}(\hat{r}_0)
}
\label{eq:greenianexp}
\end{eqnarray}
where $r<r_0<nL\neq 0$, and we have introduced the \emph{structure coefficients}
\begin{equation}
\Gamma_{lm,l^\prime m^\prime}(\veb{k},q) =
4\pi\ i^{l-l^\prime} \sum_{JK}{
i^{-J} D_{JK}(\veb{k},q)\
C_{JK;lm,l^\prime m^\prime}
}
\label{eq:structurecoeff}
\end{equation}
In the previous definition $C_{JK;lm,l^\prime m^\prime}$
are the so called Gaunt coefficients (simply related to the Wigner 3--$j$ symbols) 
defined as follows
\begin{equation}
C_{JK;lm,l^\prime m^\prime} = \int{ d\Omega_\veb{k}
Y_{JK}(\hat{k})Y_{lm}^*(\hat{k}) 
Y_{l^\prime m^\prime}(\hat{k})
}
\label{eq:W3j}
\end{equation}
while $D_{lm}(\veb{k},q)$ are the so called \emph{reduced
structure coefficients}. The time consuming part of a KKR
calculation is given by the numerical evaluation of the
reduced structure coefficients. For this reason there are
many equivalent expressions of these quantities some of
which are given in appendix~\ref{sec:app2}. Here below we limit ourself to
observe that the reduced coefficients can be written as
\begin{equation}
D_{lm}(\veb{k},q) = \frac{d_{lm}(\veb{\theta},Q)}{L}
\label{eq:dimensionlesssc}
\end{equation}
where the $d_{lm}(\veb{\theta},Q)$'s are dimensionless quantities
(see eqs.~(\ref{eq:redstrcoeffn}))
and where we have used the following definitions
\begin{equation}
\veb{\theta} = \veb{k} L \qquad\qquad\mbox{and}\qquad\qquad \veb{Q} = \frac{L\veb{q}}{2\pi}
\end{equation}
%

\subsection{Generalized L\"uscher quantization condition}

The generalized L\"uscher quantization condition, can be now easily
derived by substituting in the condition stated by eq.~(\ref{eq:nonhomsolnophiR_G})
the partial wave expansion of the wavefunctions given in eq.~(\ref{eq:psiexp})
and that of the greenian given in eq.~(\ref{eq:greenianexp}). We obtain
\begin{eqnarray}
&q\sum_{lm}&{
j_l(qr)\ Y_{lm}(\hat{r})\ \left[n_l,R_l\right]\ c_{lm}
} + 
\nonumber \\ \nonumber \\ 
&&+\sum_{lm l^\prime m^\prime}{
j_l(qr) Y_{lm}(\hat{r})\
\Gamma_{lm, l^\prime m^\prime}\
\left[j_{l^\prime},R_{l^\prime}\right] c_{l^\prime m^\prime}
} = 0 
\end{eqnarray}
that, making
use of the Wronskian relations of eq.~(\ref{eq:wronskian}),
can be rewritten as
\begin{equation}
q \cos{\delta_l(q)}\ c_{lm} + 
\sum_{l^\prime m^\prime}{
\Gamma_{lm, l^\prime m^\prime}\
\sin{\delta_{l^\prime}(q)} c_{l^\prime m^\prime}
} = 0
\end{equation}
The previous one is a linear homogeneous system of equations 
having as unknowns the coefficients $c_{lm}$.
The following compatibility determinantal equation
\begin{equation}
\det\left[
\Gamma_{lm, l^\prime m^\prime}(\veb{k},q) + 
q \delta_{l l^\prime}\delta_{m m^\prime}\ \cot{\delta_l(q)}
\right] = 0
\label{eq:kkrluscher}
\end{equation}
is the quantization conditions for the energy of the
two--particle states on a finite volume. This equation can be used
to calculate the scattering phases once the energy eigenvalues
are know (for example form lattice calculations) or vice versa
to calculate the spectrum of the two--particle state given
the scattering phases (for example by the analytical
knowledge of the interaction potential).

We want to stress again that the results obtained in this section
are valid up to exponentially small corrections. In particular
we observe that in order eq.~(\ref{eq:kkrluscher}) to be used
in quantum field theory the physical size of the finite volume
has to be large enough to exclude the presence of residual
polarization effects. These observations apply also to
the results of the following sections.

\subsection{Singular solutions}
\label{sec:singularsolutions}

In our derivation of the quantization conditions we have
assumed that the solutions of the homogeneous time--independent Shr\"odinger
equation
\begin{equation}
\left( \triangle + q^2 \right) \phi_{\veb{k}}(\veb{r}) = 0
\label{eq:homtiSE}
\end{equation}
where absent. In order $\phi_{\veb{k}}(\veb{r})$ to be 
different from zero the condition 
\begin{equation}
q =  \frac{\mo{\veb{\theta} + 2\pi\veb{n} }}{L}
\end{equation}
has to be satisfied for some integer vector $\veb{n}$ at fixed $\veb{\theta}$
together with a quantization condition that can be derived
along the same lines that we have followed to obtain eq.~(\ref{eq:kkrluscher}).
This can happen only on particular volumes $L$ and/or for
particular interactions. In the following we will not be
interested in this phenomenologically not relevant
situation and we refer the interested reader to ref.~\cite{Luscher:1990ux}
for a detailed discussion of the $\veb{\theta}=\veb{0}$ case.

\subsection{$S$--wave interactions}
\label{sec:Swave}

From the phenomenological point of view it is relevant
the situation in which all the scattering phases can be assumed to
vanish except the $S$--wave one
\begin{equation}
\delta_l(q) = 0\;, \qquad l > 0
\label{eq:alldeltazero}
\end{equation}
In this case the quantization conditions simplify extremely.
Indeed we have
\begin{equation}
C_{00;00,00} = \frac{1}{(\sqrt{4\pi})^3}\int{ d\Omega_\veb{k}
} = \frac{1}{\sqrt{4\pi}}
\end{equation}
that substituted in eq.~(\ref{eq:kkrluscher}) together
with eq.~(\ref{eq:alldeltazero}) gives us
\begin{equation}
\tan{\delta_0(q)} = -\frac{q}{\sqrt{4\pi}\ D_{00}(\veb{k},q)}
\end{equation}
or, equivalently,
\begin{equation}
\tan{\delta_0(q)} = -\frac{\sqrt{\pi}\ Q}{d_{00}(\theta,Q)}
\label{eq:swaveequation}
\end{equation}
As we have observed in eq.~(\ref{eq:dimensionlesssc}) 
(and shown in eqs.~(\ref{eq:redstrcoeffn})), the 
right hand side of the previous equation is a
dimensionless quantity that can be computed
once and forever; we set
\begin{equation}
\tan{\phi(\theta,Q)} = \frac{\sqrt{\pi}\ Q}{d_{00}(\theta,Q)}
\label{eq:tanphi}
\end{equation}
The definition of $\phi(\theta,Q)$ is completed by requiring the continuity
of this function by respect the variable $Q$ for each value of $\theta$
and by the condition
\begin{equation}
\phi(\theta,Q=0) = 0\;, \qquad 0 \le \theta_i < 2\pi
\end{equation}
%

\section{Generalized Lellouch--L\"uscher formula}
\label{sec:genLLformula}

In this section we give a generalization of the LL formula~\cite{Lellouch:2000pv}
that connects the weak matrix
element of the decay $K \rightarrow (\pi\pi)_I$ (being $I$ the
isospin of the state) calculated on a finite volume with the corresponding quantity
in the infinite volume limit up to exponentially vanishing
corrections.
In the following we work under the same hypotheses of ref.~\cite{Lellouch:2000pv},
i.e. we consider a theory of spinless pions and kaons of masses such that
the condition
\begin{equation}
2m_\pi < m_K < 4m_\pi
\end{equation}
is satisfied. We further assume that the pions scatter purely elastically
below the threshold for the production of four pions and that the kaon
is stable in the absence of weak interactions. When the weak interactions
are switched--on the kaon is allowed to decay into two pions and
the transition amplitude is given by
\begin{equation}
T(K\mapsto \pi\pi) = A(\overline{q})\ e^{i\delta_0(\overline{q})}
\label{eq:transamplitude_inf}
\end{equation}
where $A$ is real, $\delta_0$ is the $S$--wave scattering phase
of the outgoing two--pion state and $\overline{q}$ is the
pion momentum in the center-of-mass frame
\begin{equation}
\overline{q} = \frac{1}{2} \sqrt{m_K^2-4m_\pi^2}
\end{equation}
In writing eq.~(\ref{eq:transamplitude_inf}) we have
assumed the standard relativistic normalizations for the one--particle
states together with the LSZ constraints on their phases
\begin{equation}
\bra{0} \varphi_\pi(x) \ket{\pi(p)} = \sqrt{Z_\pi}\ e^{-ipx}\;,
\qquad
\bra{0} \varphi_K(x) \ket{K(p)} = \sqrt{Z_K}\ e^{-ipx}
\end{equation}
where $\varphi_\pi$ and $\varphi_K$ are the pion and kaon
interpolating fields respectively. 

Let us now consider the same theory on a finite volume of linear
extension $L$. The finite volume one--particle states
are \emph{normalized to unity} while their phases can be chosen
arbitrarily. Let us call
\begin{equation}
A_L(\overline{q}) = \bra{\pi\pi^L} H_W \ket{K^L}
\end{equation}
the transition matrix element on the finite volume with
$H_W$ the weak interactions Hamiltonian.
Our generalization of the LL formula is given by
\begin{equation}
\mo{A(\overline{q})}^2 = 8\pi \left\{
Q\frac{\partial \phi(\theta,Q)}{\partial Q}
+
q\frac{\partial \delta_0(q)}{\partial q}
\right\}_{q=\overline{q}}
\left(
\frac{m_K}{\overline{q}}
\right)^3
\mo{A_L(\overline{q})}^2
\label{eq:genLL}
\end{equation}
where $\phi(\theta,Q)$ has been defined previously in eq.~(\ref{eq:tanphi}).
We omit the derivation of this powerful relation that follows
straightforwardly by repeating the same 
arguments of ref.~\cite{Lellouch:2000pv}
and is valid under the same hypotheses that
led us in section~\ref{sec:Swave} to derive eq.~(\ref{eq:swaveequation}) 
plus two additional requirements on the outgoing
two--pion state. It has to be non degenerate
and must have the same energy
of the decaying kaon,
in order the resulting infinite volume decay amplitude
to be computed at the physical point.

In the LL derivation one has $\theta=0$ and the
two--pions state of definite isospin happens to have an energy equal to the kaon mass only on
a certain particular volume. In our generalization of their formula
the $\theta$--dependence of $\phi(\theta,Q)$ can be used in order
to obtain a sequence of volumes of growing sizes 
on which the calculations can be actually performed
(see the following section).
Successively, by studying the residual functional dependence of the results
upon the volume it will also be possible to
answer to the questions raised in ref.~\cite{Lin:2001ek} about the size of the 
finite volume corrections
coming from the presence of the inelastic threshold.

In the particular case $\theta=0$ the first seven energy levels of the outgoing two--pion
states are non degenerate, as shown in
ref.~\cite{Luscher:1990ux}. Care about the possible degeneracies of the
outgoing states has to be taken also in the case $\theta \neq 0$
for the particular choices of the Bloch's angle used in the
calculations. 

A further generalization of eq.~(\ref{eq:genLL}) by respect the
LL result can be achieved following the arguments of ref.~\cite{Lin:2001ek}
where the LL formula is shown to be valid for all the states
below the inelastic threshold and even outside the physical point ($q= \overline{q}$).

\subsection{Choice of the boundary conditions in numerical simulations}
\label{sec:bcsim}

Both the quantization conditions of eq.~(\ref{eq:kkrluscher})
and the generalized LL formula of eq.~(\ref{eq:genLL})
have been derived in the reference
frame of the two--pions center of mass. 
Although a generalization of the
quantization condition to other reference frames could be obtained
along the same lines of ref.~\cite{Rummukainen:1995vs}, 
when these formulas are used in the context of lattice QCD calculations
the $\theta$--angles corresponding to the up, down
and strange quarks have to be fixed in such a way that
the decaying kaon is at rest and that the total momentum
of the two--pion states is zero.

Let us consider, for example, the decay of a neutral
kaon in charged pions. In order to have the kaon at rest, 
considering that the anti--quarks
have a $\theta$--angle opposite to that of the corresponding quarks,
one has to chose strange and down quarks boundary conditions so that
\begin{equation}
\veb{\theta}_s = \veb{\theta}_d
\end{equation}
At the same time, the positive charged pion 
interpolating field will have an overall $\theta$--angle
given by
\begin{equation}
\veb{\theta}_{\pi^+} = \veb{\theta}_u-\veb{\theta}_d
\end{equation}
while for the negative pion one has $\veb{\theta}_{\pi^-} = -\veb{\theta}_{\pi^+}$, 
so that the total momentum of the two--pion system is zero.
There is no problem to fulfill these requirements in the
quenched theory but, when one considers the full theory,
some technical complications arise.
Indeed current simulation algorithms require the product of 
the quark determinants (that depend upon $\theta$) to be non negative and, when Wilson
fermions are considered, this implies that
\begin{equation}
\veb{\theta}_u = \veb{\theta}_d
\end{equation}
i.e. a vanishing $\theta$--angle for the interpolating
fields of the charged pions. 
Nevertheless this complication does not arise when
Ginsparg--Wilson fermions are considered because in this case
the determinant of each quark is non negative.

\section{Volume dependence of a two--pion state energy  spectrum}
\label{sec:voldep}

We are going to study the eigenvalues of a two--pion system on finite
volumes assuming that the scattering phases for $l\ge 4$ are
small in the elastic region (note that $\delta_l(q)$ is proportional to
$q^{2l+1}$ at low momenta). Under these hypotheses the energy quantization 
condition takes the simple form of eq.~(\ref{eq:swaveequation}).

\begin{figure}[t]
\begin{center}
\epsfig{file=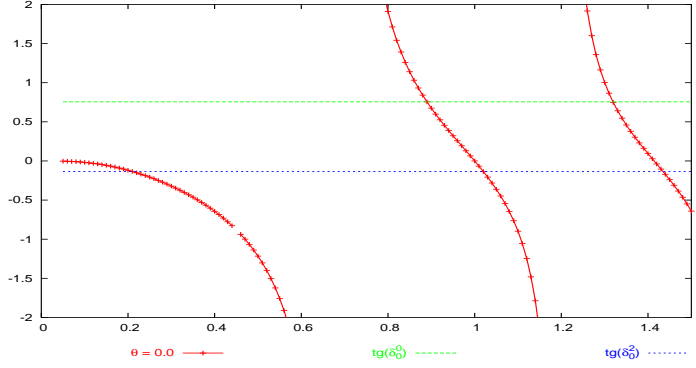,width=0.7\textwidth,height=0.35\textwidth}
\caption{The red points represent $-\tan{\phi(\theta,Q)}$ as function of $Q$
for $\theta=0.0$. The different branches are the two--particle
eigenvalues curves. The straight lines are the tangents of the scattering phases
of two pions, having isospin $I=0$ or $I=2$,
computed in chiral perturbation theory at the kaon mass.}
\label{fig:eigen1}
\end{center}
\vskip 35pt
\end{figure}
\begin{figure}[t]
\begin{center}
\epsfig{file=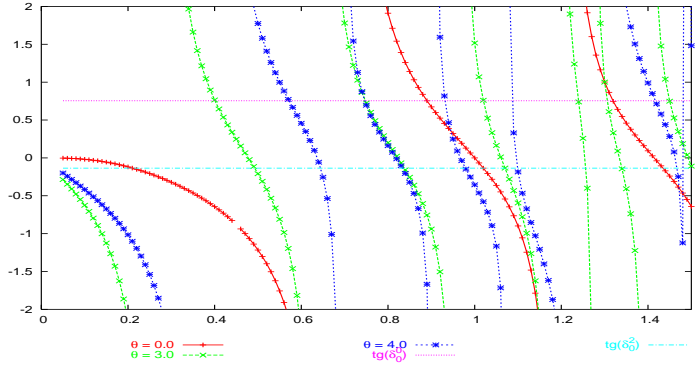,width=0.7\textwidth,height=0.35\textwidth}
\caption{The points represent $-\tan{\phi(\theta,Q)}$ as functions of $Q$
for different values of $\theta$. The case $\theta =2.5$ is not shown
to help the eye.
The straight lines are the tangents of the scattering phases
of two pions, having isospin $I=0$ or $I=2$,
computed in chiral perturbation theory at the kaon mass.}
\label{fig:eigen2}
\end{center}
\vskip 35pt
\end{figure}
In fig.~\ref{fig:eigen1} we show the opposite of~$\tan\left[\phi(\theta,Q)\right]$ for the particular
choice $\theta=0.0$. The points in which this function coincide
with the function $\tan{\delta_0(q)}$ at fixed $\theta$ and $L$ represent the two--particle
eigenvalues on the given volume but, in the following
we want to perform a slightly different 
phenomenological analysis: \\
\begin{itemize}
\item we fix the energy
  of the two particle state so that  $q=\overline{q}$\\

\item  we compute
  the corresponding scattering phase $\delta_0(\overline{q})$ 
  (using the one--loop chiral perturbation theory results
  for these 
  quantities~\cite{Gasser:1983kx,Gasser:1983yg,Gasser:1984gg,Gasser:1990ku,Knecht:1995tr})\\

\item we find the values of $\overline{Q}_n(\theta)$ such that the
  condition
\begin{equation}
\tan{\delta_0(\overline{q})} = -\tan{\phi(\theta,\overline{Q})}
\label{eq:qbartheta}
\end{equation}
is satisfied.\\

\item the volume $\overline{L}_n(\theta)$ on which
the $n^{\mbox{th}}$ two--particle eigenvalue is equal to $\overline{q}$ is computed according to
\begin{equation}
\overline{L}_n(\theta) = 2\pi \frac{\overline{Q}_n(\theta)}{\overline{q}}
\label{eq:lbar}
\end{equation}
\end{itemize}
\vskip 10pt

\noindent Following these steps we are able to find for each value of $\theta$ the finite volumes
on which a given energy level of a two--pion state equals the kaon mass.
In ref.~\cite{Lellouch:2000pv} the authors have not considered the possibility of having
$\theta \neq 0$ and their analysis, that we reproduce in fig.~\ref{fig:eigen1}, gives
the following results
\begin{eqnarray}
&&I=0 \qquad \qquad \overline{Q}_1(0) = 0.89 \qquad \qquad \overline{L}_1(0) = 5.34 \mbox{ fm}
\nonumber \\ 
&&I=2 \qquad \qquad \overline{Q}_1(0) = 1.02 \qquad \qquad \overline{L}_1(0) = 6.09 \mbox{ fm}
\end{eqnarray}
i.e. the numerical simulations should be performed on volumes of order of $5.5$~fm that,
in the case of unquenched calculations,
could be too large even for the next--generation super computers.
In fig.~\ref{fig:eigen2} we repeat the same analysis for other allowed
values of $\theta$. As can be seen a careful choice of the ``Bloch angle'' shifts the position
of the first eigenvalue curves at lower values of $Q$ and consequently, being $q$ fixed, at
lower values of the physical volume.
\begin{table}[t]
\begin{center}
\begin{tabular}{c c c c}
\toprule
\hspace{25pt} {\large \bf $\theta$}         \hspace{25pt}  & 
\hspace{25pt} {\large \bf $I$}              \hspace{25pt}  & 
\hspace{25pt} {\large \bf $\overline{Q}_1$} \hspace{25pt}  & 
\hspace{25pt} {\large \bf $L_1$}            \hspace{25pt}  \\
\toprule
0.0        &  0   &  0.890             & 5.34   \\
0.0        &  2   &  1.015             & 6.09   \\[5pt]
2.5        &  0   &  0.295             & 1.78   \\
2.5        &  2   &  0.415             & 2.49   \\[5pt]
3.0        &  0   &  0.400             & 2.40   \\
3.0        &  2   &  0.490             & 2.94   \\[5pt]
4.0        &  0   &  0.569             & 3.41   \\
4.0        &  2   &  0.645             & 3.86   \\
\bottomrule
\end{tabular}
\caption{ Volumes on which the energy of the first eigenvalue two--pion state equals
the $K$--meson mass. $I$ is the isospin of the state.
}
\label{tab:volumes}
\end{center}
\vskip 35pt
\end{table}

The resulting volumes corresponding to our particular choices of $\theta$ are
given in tab.~\ref{tab:volumes}. As can be seen the volumes to be simulated are of the
order of $3$~fm that are accessible on present super--computers.
A similar analysis can be repeated for all the allowed values of $\theta$
($0\le \theta_i < 2\pi$) in order to obtain a longer sequence of finite volumes of
growing physical sizes on which the decay amplitudes can be computed at the physical
point. As already pointed out in the previous section, the size of the
finite volume corrections to eq.~(\ref{eq:swaveequation}) and to eq.~(\ref{eq:genLL})
can be quantified by repeating the calculations on this sequence of volumes 
and by studying the functional dependence of the results upon the box extension.

\section{Conclusions}
\label{sec:conc}

In this work we have studied the spectrum of a two particle
state on a finite volume with Bloch's boundary conditions.

We have found a quantization condition that generalizes a
result previously obtained by L\"uscher in the case of periodic
boundary conditions and that relates the  
energy eigenvalues of a two--particle system on a finite volume 
with their infinite volume scattering phases. The formula
is valid up to exponentially small finite volume corrections
and allows the calculations of the scattering phases
at small physical momenta by calculating, for example on the lattice, 
the energy spectrum of the two--particle state on volumes of the order of $3$~fm.

From our quantization condition it follows
straightforwardly a generalization of the
Lellouch--L\"uscher formula that connects the
finite volume amplitudes for the decays of a kaon
into two pions with the corresponding quantities in the infinite volume.
We have argued that the decay amplitudes can be obtained
simulating on the lattice finite volumes of different
physical extensions in order to quantify the size
of the residual finite volume corrections and thus to
obtain results with a phenomenologically relevant accuracy.

\ack{ 
We warmly thank M.~L\"uscher for drawing our attention
on the subject of this work and for useful comments on the
manuscript.
We are indebted with R.~Petronzio for invaluable suggestions
and for his help during the course of the work.
Useful discussions with M.~Testa are also gratefully acknowledged.
}

\vskip 70pt
\begin{center}
{\large \bf Appendices}
\end{center}

\appendix
\section{Derivation of the partial wave expansion of the greenian}
\label{sec:app1}

In this appendix we derive the
partial wave expansion of the greenian of eq.~(\ref{eq:greenianexp})
that we have used to derive the generalized L\"uscher's quantization conditions. 
We start by recalling the well known Neumann's expansion of the Green's function
\begin{equation}
 -\frac{1}{4\pi} \frac{e^{iq\mo{\veb{r}-\veb{r}^\prime}}}{\mo{\veb{r}-\veb{r}^\prime}}
=
q\sum_{lm}j_l(qr)Y_{lm}(\hat{r})\left[
n_l(qr^\prime) - i j_l(qr^\prime)
\right] Y^*_{lm}(\hat{r}^\prime)
\end{equation}
The previous relation is valid provided that $r<r^\prime$ otherwise one has
to exchange $\veb{r}$ and $\veb{r}^\prime$ in the right--hand side.
We insert the Neumann expansion in the expression for the greenian
that we report below for clarity
\begin{equation}
g_\veb{k}(\veb{r}-\veb{r}_0;q) = 
 -\frac{1}{4\pi}
\sum_{\veb{n} \in {\bf Z}^3}{
e^{i\veb{k}\cdot\veb{n}L} 
\frac{e^{iq\mo{\veb{r}-\veb{r}_0-\veb{n}L}}}{\mo{\veb{r}-\veb{r}_0-\veb{n}L}}
}
\label{eq:greenian2}
\end{equation}
For convenience we separate out the term with $\veb{n} = \veb{0}$ from the remaining
terms and we use the Neumann expansion identifying $\veb{r}^\prime$ with $\veb{r}_0$ in the first
case and with  $\veb{n}L$ in the second case.
We end up with
\begin{eqnarray}
&g_\veb{k}(&\veb{r}-\veb{r}_0;q) =
\nonumber \\ \nonumber \\
&&=-\frac{1}{4\pi} \frac{\cos\mo{\veb{r}-\veb{r}_0}}{\mo{\veb{r}-\veb{r}_0}} +
\sum_{lm}{
D_{lm}(\veb{k},q)\ j_l(q\mo{\veb{r}-\veb{r}_0})\ Y_{lm}(\widehat{\veb{r}-\veb{r}_0})
}
\label{eq:identity}
\end{eqnarray}
where the so called reduced structure coefficients are given by
\begin{eqnarray}
&D_{lm}&(\veb{k},q) = 
\nonumber \\\nonumber \\
&&=\sum_{\veb{n} \in {\bf Z}^3-\left\{\veb{0}\right\}}{
e^{i\veb{k}\cdot\veb{n}L} Y_{lm}(\hat{n})
\left[
n_l(qnL) - i j_l(qnL)
\right]
}\, - i \frac{q}{\sqrt{4\pi}}\delta_{l0}\delta_{m0}
\label{eq:redstrcoeff}
\end{eqnarray}
There are many different, although equivalent from the
mathematical point of view, ways to express the
reduced structure coefficients some of which are more
convenient than the previous relation for a numerical
computation of these quantities. In the appendix~\ref{sec:app2}
we derive an expression suitable for the numerical calculation.

We need to recall another well known identity
that can be easily proved by using the expansion in spherical
harmonics of a plane wave (see eq.~(\ref{eq:planespherical}))
\begin{eqnarray}
&i^{JK} j_J(qR) &Y_{JK}(\hat{R}) =
\nonumber \\\nonumber \\
&&=4\pi \sum_{lm,l^\prime m^\prime}{
i^{l-l^\prime} C_{JK;lm,l^\prime m^\prime}\ j_l(qr)\ j_{l^\prime}(qr_0)\ 
Y_{lm}(\hat{r})  Y_{l^\prime m^\prime}(\hat{r}_0) 
}
\label{eq:identity2}
\end{eqnarray}
where we have called $R = \mo{\veb{r}-\veb{r}_0}$ and we have introduced
the Gaunt coefficients (see eq.~(\ref{eq:W3j}) in the text)
\begin{equation}
C_{JK;lm,l^\prime m^\prime} = \int{ d\Omega_\veb{k}
Y_{JK}(\hat{k})Y_{lm}^*(\hat{k}) 
Y_{l^\prime m^\prime}(\hat{k})
}
\end{equation}
Inserting the identity of eq.~(\ref{eq:identity2}) in the partial wave
expansion of the greenian as given in eq.~(\ref{eq:identity})
we are able to rewrite this expansion in the same form of 
eq.~(\ref{eq:greenianexp}) that we have used in the text to derive the
quantization conditions, i.e.
\begin{eqnarray}
g_\veb{k}(\veb{r}-\veb{r}_0;q) &=&
q\sum_{lm}{
j_l(qr)\ Y_{lm}(\hat{r})
n_l(qr_0)\ Y_{lm}^*(\hat{r}_0)
}
\nonumber \\ \nonumber \\
&+&
\sum_{lm l^\prime m^\prime}{
j_l(qr)\ Y_{lm}(\hat{r})\
\Gamma_{lm,l^\prime m^\prime}(\veb{k},q)\
j_{l^\prime}(qr_0)\ Y_{l^\prime m^\prime}(\hat{r}_0)
}
\label{eq:greenianexp_a}
\end{eqnarray}
where $r<r_0<nL\neq 0$, and we have introduced the structure coefficients
\begin{equation}
\Gamma_{lm,l^\prime m^\prime}(\veb{k},q) =
4\pi\ i^{l-l^\prime} \sum_{JK}{
i^{-J} D_{JK}(\veb{k},q)\
C_{JK;lm,l^\prime m^\prime}
}
\label{eq:structurecoeff_a}
\end{equation}
%

\section{Structure coefficients calculation}
\label{sec:app2}

In this appendix we derive a mathematical expression 
useful for the numerical calculation of the structure coefficients.
From eq.~(\ref{eq:structurecoeff_a}) we realize that the non trivial
part of this problem consists in the calculation of the reduced structure
coefficients. In the following we are going to derive an expression of
the reduced structure coefficients, 
different from that already given in eq.~(\ref{eq:redstrcoeff}), in order to
handle a formula suitable for the numerical evaluation.
We start again from the greenian definition
\begin{equation}
g_\veb{k}(\veb{R};q) = -\frac{1}{4\pi}\sum_{\veb{n} \in {\bf Z}^3}{
e^{i\veb{k}\cdot\veb{n}L} 
\frac{e^{iq\mo{\veb{R}- \veb{n}L}}}{\mo{\veb{R} - \veb{n}L}}
}
\label{eq:greenian31}
\end{equation}
having the aim to express it in the reciprocal space.
To this end we recall the following well known identities
\begin{eqnarray}
&& \frac{1}{(2\pi)^3} \int{ d\veb{x}\
e^{i(\veb{p}_1 - \veb{p}_2)\cdot \veb{x}}
} = \delta(\veb{p}_1 - \veb{p}_2)
\\ \nonumber \\
&& \frac{1}{L^3} \sum_{\veb{p}_m}{
e^{i\veb{p}_m \cdot \veb{x}}
} = \sum_{\veb{n} \in {\bf Z}^3}{
\delta(\veb{x} - \veb{n}L)\ , \qquad \veb{p}_m = \frac{2\pi}{L}\veb{m}
}
\\ \nonumber \\
&& \lim_{\varepsilon \rightarrow 0^+}{
\frac{1}{(2\pi)^3} \int{ d\veb{p}\
\frac{e^{i\veb{p}\cdot \veb{x}}}{p^2 -(q^2+i\varepsilon)}
}
} = \frac{1}{(4\pi)} \frac{e^{iqx}}{x}
\end{eqnarray}
that allow us to write the following chain of equalities
\begin{eqnarray}
g_\veb{k}(\veb{R};q) &=& - \frac{1}{(2\pi)^3} \lim_{\varepsilon \rightarrow 0^+}{
\sum_{\veb{n} \in {\bf Z}^3}{
e^{i\veb{k}\cdot\veb{n}L}\
\int{ d\veb{p}\
\frac{e^{i\veb{p}\cdot (\veb{R}-\veb{n}L)}}{p^2 -(q^2+i\varepsilon)}
}
}
}
\nonumber \\ \nonumber \\
&=& - \frac{1}{(2\pi)^3} \lim_{\varepsilon \rightarrow 0^+}{
\int{ d\veb{p}\
\frac{e^{i\veb{p}\cdot \veb{R}}}{p^2 -(q^2+i\varepsilon)}\
\sum_{\veb{n} \in {\bf Z}^3}{
e^{i(\veb{k}-\veb{p})\cdot\veb{n}L}
}
}
}
\nonumber \\ \nonumber \\
&=& -\lim_{\varepsilon \rightarrow 0^+}{
\int{ d\veb{p}\
\frac{e^{i\veb{p}\cdot \veb{R}}}{p^2 -(q^2+i\varepsilon)}\
\frac{1}{(2\pi)^3} 
\int{d\veb{x} 
\sum_{\veb{n} \in {\bf Z}^3}{ \delta(\veb{x}-\veb{n}L)
e^{i(\veb{k}-\veb{p})\cdot\veb{x}}
}
}
}
}
\nonumber \\ \nonumber \\
&=& -\lim_{\varepsilon \rightarrow 0^+}{
\frac{1}{L^3} 
\sum_{\veb{p}_m}{
\int{ d\veb{p}\
\frac{e^{i\veb{p}\cdot \veb{R}}}{p^2 -(q^2+i\varepsilon)}\
\frac{1}{(2\pi)^3} \int{d\veb{x} 
e^{i(\veb{k}+\veb{p}_m-\veb{p})\cdot\veb{x}}
}
}
}
}
\nonumber \\ \nonumber \\
&=& -\lim_{\varepsilon \rightarrow 0^+}{
\frac{1}{L^3} 
\sum_{\veb{p}_m}{
\int{ d\veb{p}\
\frac{e^{i\veb{p}\cdot \veb{R}}}{p^2 -(q^2+i\varepsilon)}\
\delta(\veb{k}+\veb{p}_m-\veb{p})
}
}
}
\nonumber \\ \nonumber \\
&=& -\frac{1}{L^3} 
\sum_{\veb{p}_m}{
\frac{e^{i(\veb{k}+\veb{p}_m)\cdot \veb{R}}}{(\veb{k}+\veb{p}_m)^2 - q^2}
}
\end{eqnarray}
i.e., the greenian expression in the reciprocal space is given by
\begin{equation}
g_\veb{k}(\veb{R};q) 
= -\frac{1}{L^3} 
\sum_{\veb{q}_m}{
\frac{e^{i\veb{q}_m\cdot \veb{R}}}{\veb{q}_m^2 - q^2}\ , \qquad \veb{q}_m = \veb{k}+\frac{2\pi}{L}\veb{m}
}
\end{equation}
We are now ready to derive an expression for the KKR reduced structure
coefficients in the reciprocal space. To this end we need to recall
the identity expressing a plane wave as an expansion in spherical
harmonics
\begin{equation}
e^{i\veb{q}_m\cdot\veb{R}} = 4\pi\sum_{lm}{
i^{l} j_l(q_m R)\ Y_{lm}^*(\hat{q}_m)\
Y_{lm}(\hat{R})
}
\label{eq:planespherical}
\end{equation}
so that the greenian can be expanded as follows
\begin{equation}
g_\veb{k}(\veb{R};q) 
= \sum_{lm}{
\left(
-\frac{4\pi\ i^l}{L^3} 
\sum_{\veb{q}_m}{
\frac{j_l(q_m R)}{j_l(qR)}\
\frac{Y_{lm}^*(\hat{q}_m)}{\veb{q}_m^2 - q^2}
}
\right)
j_l(qR)\ Y_{lm}(\hat{R})
}
\label{eq:greenian3}
\end{equation}
In order to reproduce the same structure of eq.~(\ref{eq:identity})
we make the following observation
\begin{eqnarray}
g_\veb{k}(\veb{R};q) &=& g_\veb{k}(\veb{R};q) -\frac{1}{4\pi} \frac{cos(qR)}{R} 
+\frac{1}{4\pi} \frac{cos(qR)}{R} 
\nonumber \\\nonumber \\
&=& g_\veb{k}(\veb{R};q) -\frac{1}{4\pi} \frac{cos(qR)}{R} 
+\frac{q}{4\pi} \cot(qR)\ \frac{sin(qR)}{R} 
\nonumber \\\nonumber \\
&=& g_\veb{k}(\veb{R};q) -\frac{1}{4\pi} \frac{cos(qR)}{R} 
+\frac{q}{4\pi} \cot(qR)\ j_0(qR) 
\end{eqnarray}
inserting the last identity in eq.~(\ref{eq:greenian3}) we end up
with
\begin{equation}
g_\veb{k}(\veb{R};q) 
= -\frac{1}{4\pi} \frac{cos(qR)}{R} + \sum_{lm}{
D_{lm}(\veb{k},q;R)
j_l(qR)\ Y_{lm}(\hat{R})
}
\label{eq:greenian4}
\end{equation}
where we have obtained
\begin{equation}
D_{lm}(\veb{k},q;R) =
-\frac{4\pi i^l}{L^3} 
\sum_{\veb{q}_m}{
\frac{j_l(q_m R)}{j_l(qR)}\
\frac{Y_{lm}^*(\hat{q}_m)}{\veb{q}_m^2 - q^2}
}
+\frac{q}{\sqrt{4\pi}} \cot(qR)\ \delta_{l0}\ \delta_{m0}
\end{equation}
This expression requires some comments. First we want to note
that
\begin{equation}
Y_{lm}^*(\hat{q}_m) = Y_{lm}(\hat{q}^\prime_m) 
\qquad \veb{q}_m = (q_m^1,q_m^2,q_m^3)
\qquad \veb{q}^\prime_m = (q_m^1,-q_m^2,q_m^3)
\end{equation}
so that, being the rest of the sum argument a function of the 
modulus of $\veb{q_m}$ only, we can write
\begin{equation}
D_{lm}(\veb{k},q;R) =
-\frac{4\pi i^l}{L^3} 
\sum_{\veb{q}_m}{
\frac{j_l(q_m R)}{j_l(qR)}\
\frac{Y_{lm}(\hat{q}_m)}{\veb{q}_m^2 - q^2}
}
+\frac{q}{\sqrt{4\pi}} \cot(qR)\ \delta_{l0}\ \delta_{m0} 
\label{eq:redcoeff1}
\end{equation}
The second observation concerns the $R$--dependence of
the reduced structure coefficients. Indeed, by the comparison
of eq.~(\ref{eq:redstrcoeff}) with the previous equation
we learn that this functional dependence is fictitious. This can be
also understood by a careful analysis of eq.~(\ref{eq:greenian4}).
We know that the term
\begin{equation}
n_0(qR) = \frac{cos(qR)}{qR} 
\end{equation}
satisfies by its own the equation
\begin{equation}
\left( \triangle + q^2 \right) n_0(qR) = -\frac{4\pi}{q}\delta(\veb{R}) 
\end{equation}
so that the remaining terms of the sum in eq.~(\ref{eq:greenian4}) must be
regular solutions of the homogeneous Helmholtz's equation. A general solution
of this kind can be expressed as a linear combination of the form
\begin{equation}
\hat{g}_\veb{k}(\veb{R};q) 
= \sum_{lm}{
D_{lm}(\veb{k},q)
j_l(qR)\ Y_{lm}(\hat{R})
}
\end{equation}
where the $D_{lm}(\veb{k},q)$ are constants by respect to $R$.
We also learn from textbooks on mathematical functions that
near the origin the spherical Bessel functions deal as
\begin{equation}
j_l(qR) \stackrel{R\rightarrow 0}{\longrightarrow} (qR)^l 
\end{equation}
The fictitious $R$--dependence of the reduced structure coefficients
as given in eq.~(\ref{eq:redcoeff1}) can be eliminated taking the limit
for $R$ that goes to zero and using the previous equation.

One has to consider eq.~(\ref{eq:redcoeff1}) as a properly regulated form of
the KKR reduced structure coefficients and think to the following
\begin{equation}
D_{lm}(\veb{k},q) =
-\frac{4\pi i^l}{q^l L^3} 
\sum_{\veb{q}_m}{
\frac{q_m^l\ Y_{lm}(\hat{q}_m)}{\veb{q}_m^2 - q^2}
}
+\frac{\delta_{l0}\ \delta_{m0} }{\sqrt{4\pi}} \lim_{R\rightarrow 0}{\frac{1}{R}}
\label{eq:redcoeff_final}
\end{equation}
as a formal expression of the same objects.

\subsection{Ewald's sums}

In ref.~\cite{Kohn:1954} Kohn and Rostoker considered a method
particularly convenient from the numerical point of view to
evaluate the reduced structure coefficients. They pointed out
that, following a prescription due to Ewald~\cite{Ewald:1921},
the sum
\begin{equation}
S(x) = \sum_{\veb{q}_m}{
\frac{j_l(q_mR)\ Y_{lm}(\hat{q}_m)}{q_m^2-q_2}\
e^{\frac{q^2-q_m^2}{x}}
}
\end{equation}
approximate the needed result $S(\infty)$ with an exponential vanishing error.
In ref.~\cite{Ham:1961} the original observation of Kohn and Rostoker
was further refined by Ham and Segall. They first recall two identities
both due to Ewald; the first one is
\begin{equation}
\frac{e^{iq\mo{\veb{R}-\veb{n}L}}}{\mo{\veb{R}-\veb{n}L}} = 
\lim_{\varepsilon \rightarrow 0}{
\frac{2}{\sqrt{\pi}}\int_0^\infty{dx\ 
e^{-(\veb{R}-\veb{n}L)^2 x^2 + \frac{q^2+i\varepsilon}{4x^2}}
}}
\label{eq:ewald1}
\end{equation}
The second identity can be proved as done before to express the greenian
in the reciprocal space starting from its expression in the direct space
and states
\begin{equation}
\sum_{\veb{n} \in {\bf Z}^3}{
e^{-(\veb{R}-\veb{n}L)^2 x^2 + i\veb{k} \cdot (\veb{n}L-\veb{R})}
=
\frac{(\sqrt{\pi})^3}{(Lx)^3} \sum_{\veb{p}_m}{
e^{i\veb{p}_m\cdot \veb{R}-\frac{q_m^2}{4x^2}}
}
}\qquad \veb{p}_m = \frac{2\pi}{L}\veb{m}
\label{eq:ewald2}
\end{equation}
We now start again from the expression of the 
greenian in the direct space as given in eq.~(\ref{eq:greenian31})
and write, by using the first identity~(\ref{eq:ewald1}),
\begin{equation}
g_\veb{k}(\veb{R};q) = 
-\frac{1}{2(\sqrt{\pi})^3}\lim_{\varepsilon \rightarrow 0}{
\sum_{\veb{n} \in {\bf Z}^3}{
\int_0^\infty{dx\ 
e^{i\veb{k}\cdot \veb{n}L -(\veb{R}-\veb{n}L)^2 x^2 + \frac{q^2+i\varepsilon}{4x^2}}
}
}
}
\end{equation}
then we split the integration domain in $\left[0,\frac{\sqrt{\eta}}{2}\right]$ and
$\left[\frac{\sqrt{\eta}}{2},\infty\right]$, where $\eta$ is a positive arbitrary
constant. The greenian can be thus re-expressed as the sum of two terms
\begin{equation}
g_\veb{k}(\veb{R};q) = 
g_\veb{k}^1(\veb{R};q) + g_\veb{k}^2(\veb{R};q)  
\end{equation}
where, performing the integration in the first term and using
the identity of eq.~(\ref{eq:ewald2}) in the second one, we obtain
\begin{eqnarray}
g_\veb{k}^1(\veb{R};q) &=&
\frac{1}{L^3}\sum_{\veb{q}_m}{
\frac{e^{i\veb{q}_m\cdot \veb{R} + \frac{q^2-q_m^2}{\eta}}}{q^2-q_m^2}
}
\nonumber \\\nonumber \\
g_\veb{k}^2(\veb{R};q) &=&-\frac{1}{2(\sqrt{\pi})^3}
\sum_{\veb{n} \in {\bf Z}^3}{
\int_{\frac{\sqrt{\eta}}{2}}^\infty{dx\ 
e^{i\veb{k}\cdot \veb{n}L -(\veb{R}-\veb{n}L)^2 x^2 + \frac{q^2}{4x^2}}
}
}
\end{eqnarray}
These series are absolutely convergent for any finite $\eta>0$ and each
term is an analytic function of $q$ throughout the complex plane except at
the simple pole $q^2=q_m^2$. Expanding termwise both $g_\veb{k}^1$ and $g_\veb{k}^2$
in spherical harmonics by respect to $\veb{R}$, taking the limit $R\rightarrow 0$, and
comparing the result with the definition of the reduced structure coefficients, we find
\begin{equation}
D_{lm}(\veb{k},q) = D_{lm}^1(\veb{k},q) + D_{lm}^2(\veb{k},q) + D_{lm}^3(\veb{k},q)
\end{equation}
where
\begin{eqnarray}
D_{lm}^1(\veb{k},q) &=& \frac{4\pi}{L^3 q^l}\ e^{\frac{q^2}{\eta}}\
\sum_{\veb{q}_m}{
\frac{q_m^l}{q^2-q_m^2}\ Y_{lm}(\hat{q}_m)\ e^{-\frac{q_m^2}{\eta}}
}
\nonumber \\ \nonumber \\
D_{lm}^2(\veb{k},q) &=& -\frac{2^{l+1} L^l i^l}{q^l \sqrt{\pi}}
\sum_{\veb{n} \in {\bf Z}^3-\left\{ \veb{0}\right\}}{
n^l\ e^{i\veb{k}\cdot \veb{n}L}\ Y_{lm}(\hat{n})\
\int_{\frac{\sqrt{\eta}}{2}}^\infty{dx\ 
x^{2l}\ e^{-(x\veb{n}L)^2 + \frac{q^2}{4x^2}}
}
}
\nonumber \\ \nonumber \\ 
D_{lm}^3(\veb{k},q) &=& 
-\delta_{l0}\ \delta_{m0}\ \frac{\sqrt{\eta}}{2\pi}\sum_{s=0}^\infty{
\frac{q^{2s}}{\eta^s}\ \frac{1}{s!(2s-1)}
}
\label{eq:rednovolume_a}
\end{eqnarray}
In order to make more explicit the dependence of the reduced structure
coefficients upon the volume let us define
\begin{eqnarray}
d_{lm}(\theta,Q) &=& L D_{lm}(\veb{k},q) 
\nonumber \\ \nonumber \\ 
&=& d_{lm}^1(\theta,Q) + d_{lm}^2(\theta,Q) + d_{lm}^3(\theta,Q)
\end{eqnarray}
and rewrite eqs.~(\ref{eq:rednovolume_a}) as follows
\begin{eqnarray}
d_{lm}^1(\theta,Q) &=& \frac{i^l}{\pi Q^l}\ e^{\frac{Q^2}{\eta^\prime}}\
\sum_{\veb{Q}_m}{
\frac{Q_m^l\  Y_{lm}(\hat{Q}_m)}{Q^2-Q_m^2}\ e^{-\frac{Q_m^2}{\eta^\prime}}
}
\nonumber \\ \nonumber \\
d_{lm}^2(\theta,Q) &=& -\frac{4^{l+1} i^l \sqrt{\pi}}{Q^l}
\sum_{\veb{n} \in {\bf Z}^3-\left\{ \veb{0}\right\}}{
(\pi n)^l e^{i\veb{\theta}\cdot \veb{n}} Y_{lm}(\hat{n})
\int_{\frac{\sqrt{\eta^\prime}}{2}}^\infty{dx 
x^{2l} e^{-(2\pi x\veb{n})^2+\frac{Q^2}{4x^2}}
}
}
\nonumber \\ \nonumber \\ 
d_{lm}^3(\theta,Q) &=& 
-\delta_{l0}\ \delta_{m0}\ \sqrt{\eta^\prime}\sum_{s=0}^\infty{
\frac{Q^{2s}}{{\eta^\prime}^s}\ \frac{1}{s!(2s-1)}
}
\label{eq:redstrcoeffn}
\end{eqnarray}
where we have substituted
\begin{eqnarray}
&Q = \frac{Lq}{2\pi} \qquad \qquad \qquad 
&Q_m = \mo{ \veb{m}+ \frac{\veb{\theta}}{2\pi}}
\\
\nonumber \\
&\veb{k} = \frac{\veb{\theta}}{L} \qquad \qquad \qquad 
&\eta^\prime = \left(\frac{L}{2\pi}\right)^2\eta
\end{eqnarray}
%

\subsection{Incomplete Gamma Function}

The computation of the reduced structure coefficients is complicated
from the numerical point of view by the presence of an integral
in the definition of $d_{lm}^2(\theta,Q)$. In order to simplify
the numerical task let us recall the definition of the
\emph{incomplete gamma function}:
\begin{equation}
\Gamma(\alpha,y) = \int_y^\infty{dx\ x^{\alpha-1} e^{-x}}
\end{equation}
using the incomplete gamma function, $d_{lm}^2(\theta,Q)$ can be
rewritten in the following form
\begin{eqnarray}
&&d_{lm}^2(\theta,Q) = -\frac{i^l \sqrt{\pi}}{Q^l} \times
\nonumber \\ \nonumber \\ 
&&\times \sum_p{
\frac{Q^{2p}}{p!}
\sum_{\veb{n} \in {\bf Z}^3-\left\{ \veb{0}\right\}}{
\frac{e^{i\veb{\theta}\cdot \veb{n}}}{(\pi n)^{l-2p+1}}\ Y_{lm}(\hat{n})\
\Gamma\left( l-p+\frac{1}{2}, (\pi n)^2 \eta^\prime\right)
}}
\label{eq:D2smartL}
\end{eqnarray}
The great advantage of this expression with respect to the one given in eq.~(\ref{eq:redstrcoeffn})
is that the incomplete gamma function can be computed numerically
using a continued fraction representation:
\begin{equation}
\Gamma(\alpha,y) = \frac{e^{-y}\ y^\alpha}{
x \; + \; \frac{1-\alpha}{
1\; + \; \frac{1}{
x \; + \; \frac{2-\alpha}{
1\; + \; \frac{2}{
x \; + \; \frac{3-\alpha}{
1\; + \; \dots
}
}
}
}
}
}
\end{equation}
%


\bibliographystyle{h-elsevier} 
\bibliography{pipi}

\end{document}